# Personality cannot be predicted from the power of resting state EEG

Kristjan Korjus*[1], Andero Uusberg[2], Helen Uibo[2,3], Nele Kuldkepp[2,3], Kairi Kreegipuu[2], Jüri Allik[2,4], Raul Vicente[1], Jaan Aru[1,5]

1. University of Tartu, Institute of Computer Science, Liivi 2, 50409 Tartu, Estonia
2. University of Tartu, Institute of Psychology, Näituse 2, 50409, Tartu, Estonia
3. University of Tartu, Doctoral School of Behavioural, Social and Health Sciences, Ülikooli 18, 50090, Tartu, Estonia
4. Estonian Academy of Sciences, Kohtu 6, 10130, Tallinn, Estonia
5. University of Tartu, Institute of Public Law, Näituse 20, 50409, Tartu, Estonia

**\*Correspondence:**
Mr Kristjan Korjus
Institute of Computer Science
University of Tartu
Liivi 2-015
Tartu, 50409, Estonia
korjus@gmail.com

## Abstract

In the present study we asked whether it is possible to decode personality traits from resting state EEG data. EEG was recorded from a large sample of subjects ($n = 309$) who had answered questionnaires measuring personality trait scores of the 5 dimensions as well as the 10 subordinate aspects of the Big Five. Machine learning algorithms were used to build a classifier to predict each personality trait from power spectra of the resting state EEG data. The results indicate that the five dimensions as well as their subordinate aspects could not be predicted from the resting state EEG data. Finally, to demonstrate that this result is not due to systematic algorithmic or implementation mistakes the same methods were used to successfully classify whether the subject had eyes open or eyes closed and whether the subject was male or female. These results indicate that the extraction of personality traits from the power spectra of resting state EEG is extremely noisy, if possible at all.

## 1. Introduction

Personality can be defined as a relatively stable pattern on thinking, feeling and acting. These patterns can be explained by the idea of personality traits – underlying mechanisms that cause variation in observable personality characteristics (Deary, 2009). According to a dominant Five Factor model (FFM), observable personality is mostly determined by five major traits – Neuroticism, Extraversion, Openness, Agreeableness and Conscientiousness (McCrae & Costa, 2008; McCrae & John, 1992). Their relatively high cultural universality, temporal stability, and heritability suggest that the Big Five traits may represent some parameters of fairly specific brain networks (Corr, 2004; DeYoung & Gray, 2009; Kennis, Rademaker, & Geuze, 2012).

If traits indeed reflect individual differences in tonic brain function, then measures of baseline brain activity may provide a direct way for personality assessment. A relatively cost-effective way for quantifying the biological origins of traits could thus be developed by finding reliable correlates of trait levels from the resting state EEG signal. However, existing attempts to do this have generally yielded mixed results. For instance, an early hypothesis relating Extraversion to baseline brain arousal turned out to be a gross over-simplification (Stelmack, 1990). Another influential idea linking anterior asymmetry in EEG alpha (8-12 Hz) band power to individual differences in approach and avoidance systems of the brain (Coan & Allen, 2002; Davidson 2001), has similarly not been confirmed using meta-analytic methods (Wacker, Chavanon, & Stemmler, 2010).

On the other hand, new findings keep suggesting novel candidate parameters of resting state EEG as potential correlates of certain personality traits. For instance, mid-frontal theta power has been found to co-vary with Extraversion (Knyazev, 2009; Wacker et al., 2010). Meanwhile, the extent of negative relationships between power in lower (delta and theta) and higher (alpha and beta) frequency bands (i.e. cross-frequency anti-coupling) seems also to index approach-avoidance related individual differences (Schutter & Knyazev, 2012). Then again, this hopeful state of affairs might simply reflect the fact that the proposed correlates are still novel and negative findings remain to be published.

Most of the existing research on resting state EEG correlates of personality traits has been conducted in a hypothesis-driven way, concentrating usually on a single parameter at a time. An alternative approach would be to use data-driven techniques to first of all assess the extent to which resting state EEG signal contains information on personality and then search for relevant correlates in a more comprehensive and systematic manner. The main aim of the present study is to test such an approach. To that end we used classifiers, mathematical models that map input data to a set of classes or labels, to predict personality traits from resting state EEG signals. The classifiers were first trained using a set of data with known classes and then their performance was evaluated on data not used for the training phase.

A secondary aim of the present study is to investigate which level of personality trait description is best suited for relating to resting state EEG. In an influential paper, DeYoung and colleagues identified two lower-order aspects for each of the Big Five traits (DeYoung, Quilty, & Peterson, 2007). Subsequent research has demonstrated that this level of trait descriptions may have more homogenous brain origins (e.g. DeYoung, 2013). We therefore test if the information contained within resting state EEG is more reliably related to the 10 aspects compared to the 5 traits of the Big Five.

## 2. Results

We analyzed a large dataset collected from 309 participants. This dataset consisted of eyes open and eyes closed resting state EEG recordings (32 active electrodes) together with Big Five personality scores assessed using validated self-report questionnaires.

In the first part of the analysis, we trained statistical classifiers to map the features of the resting state EEG to personality scores of individual subjects. In particular, we used the power spectra of the EEG signals as the basis for the features or explanatory variables. The predicted variable consisted of the binarization (using a median split) of the score for each personality trait (see Methods, *Personality measures*). Given the exploratory nature of the analysis we scanned different combinations of classifier parameters and features from the EEG power spectra to find the configuration that best classified each personality trait. To avoid *cherry-picking* or over-fitting the results we always assessed the selected classifiers on a separate subset of subjects.

Thus, we used a nested cross-validation approach. We applied 10-fold cross validation and with 90% of data we used again 10-fold cross-validation in order to choose the hyper-parameters of the classifier (including different data pre-processing options, dimensionality reduction, and the choice between linear and non-linear SVM classifier, see Methods). Then, we selected the classifier that performed the best (by minimizing misclassification rate) for each personality trait. Subsequently, the remaining 10% of subjects from the first cross-validation were used to estimate the misclassification rate of the selected classifier. This procedure was repeated 10 times and the final misclassification rates represent the averages over these 10 partitions (see Methods, *Classification of the Personality Traits*).

The results for the binary classification of the test subjects are shown on Figure 1. They indicate that none of the personality traits were correctly classified above the level of significance (binomial test, $p > 0.05$) from any of the explored combinations of resting state EEG features.

Although the binarization of the personality scores is conceptually simple, robust and easy to interpret, it also leads to a loss of information that might hamper the predictive power of the classifiers. Thus, we next tried to predict the personality score of each trait as a continuous variable. For this case, instead of SVM classifiers, we used LASSO and elastic nets models of regressions. These models of regression include a penalty term that reduces the number of explanatory variables used while minimizing the error of the regression. Given the large number of features considered in our analysis, using sparse regression models allows us to control for over-fitting errors, as sparse models often tend to generalize better for novel data. Following the same nested cross-validation approach as in the binary classifier pipeline, the results showed that none of the mean squared errors was significantly better (all $p > 0.4$) than the null model hypothesis that the best prediction is the mean of the personality scores. This result indicates that essentially no additional information could be predicted from the EEG power spectrum features by using this approach.

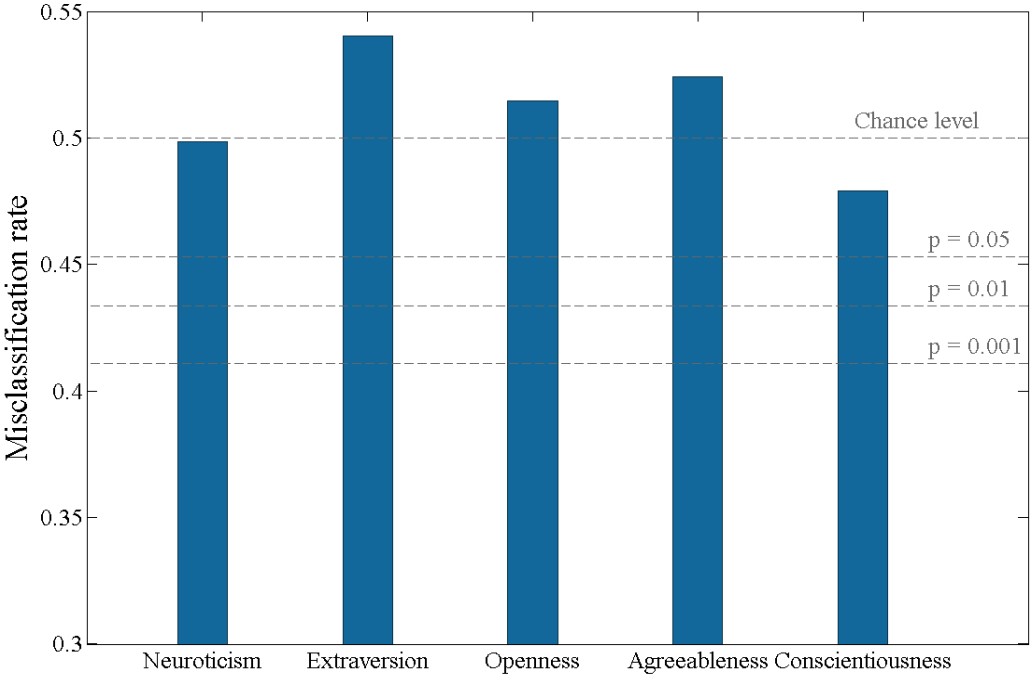

**Figure 1:** Misclassification rates of Big Five personality traits. The personality scores have been binarized with a median-split. None of the misclassification rates of five personality traits is statistically significant ($p > 0.05$).

In addition, to test an effect of age and gender related systematic personality variability, all trait scores were normalized in relation to age- and gender-specific means and standard deviations (see Methods, *Personality Measures*). The repetition of the binary classification analysis with these normalized scores still provided misclassification errors that were not statistically significant (all $p > 0.1$).

Given the possibly enhanced homogeneity of the brain origins of lower-level aspects of personality traits, we next attempted to predict the scores for the 10 lower-order aspects of the Big Five. We again used the binary classifier with the similar pipeline as for the five superordinate traits. The results are shown in Figure 2. As can be seen from Figure 2, one aspect, *Politeness,* is statistically significant at the uncorrected level. However, after the false discovery rate correction none of the *p* values remain significant. Therefore, we could not classify the lower-order aspects of Big Five personality traits from the power spectra of resting state EEG.

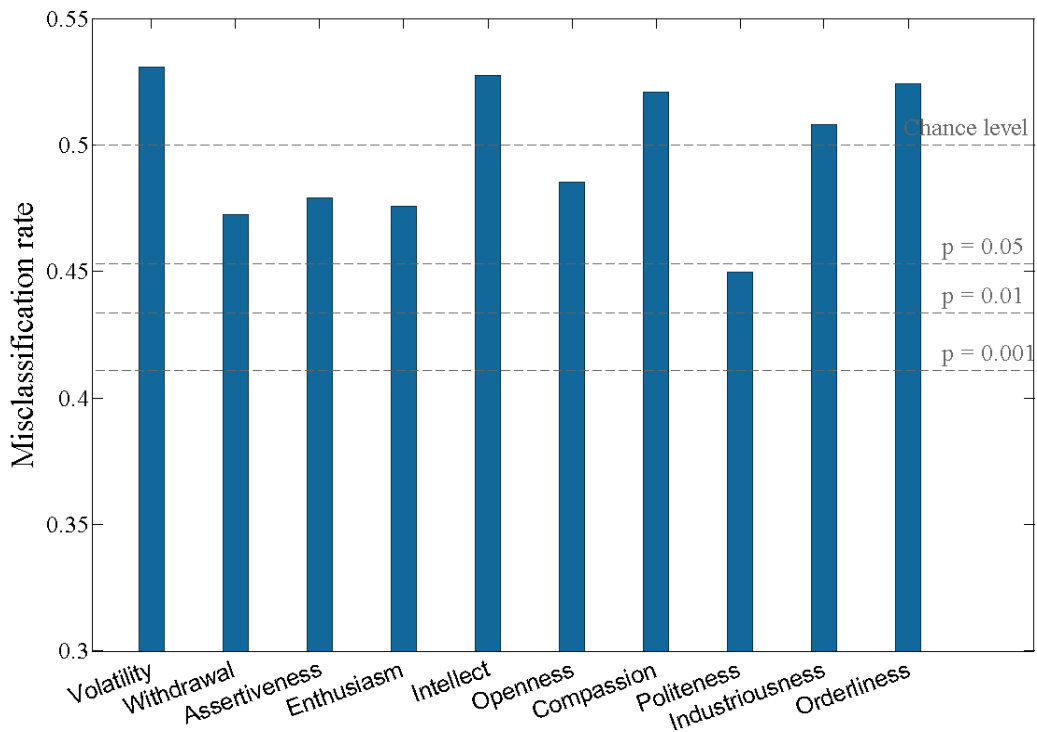

**Figure 2:** Misclassification rates of lower-order aspects of Big Five personality traits. The aspect scores have been binarized with a median-split. Although the significance of the misclassification rate for *Politeness* is below $p < 0.05$, it does not remain significant after false discovery rate correction.

To test the general validity of our analysis pipeline, predicting was also performed in a situation where clear classifiable information was present in the dataset. To that end we tried to classify whether the eyes of the subject were open or closed and also whether the subject was female or male.

Firstly, the subjects were assigned into two classes: for half of the subjects eyes open data were used and for another half eyes closed data were used. Using the same pipeline with nested cross-validation of 648 models on 309 subjects, the achieved classification rate of the classifier was 81% (misclassification rate of 19%, $p \ll 0.001$). Given that we did not control for the gender and age variability of the subjects, this result indicates that when a clear pattern of information was present in the data, our algorithm was able to extract it and perform almost optimally.

Secondly, gender of all the 309 subjects (209 female) was classified using the same nested cross-validation pipeline. Misclassification rate of 22% was achieved (chance level was 32.4% and $p = 0.00004$).

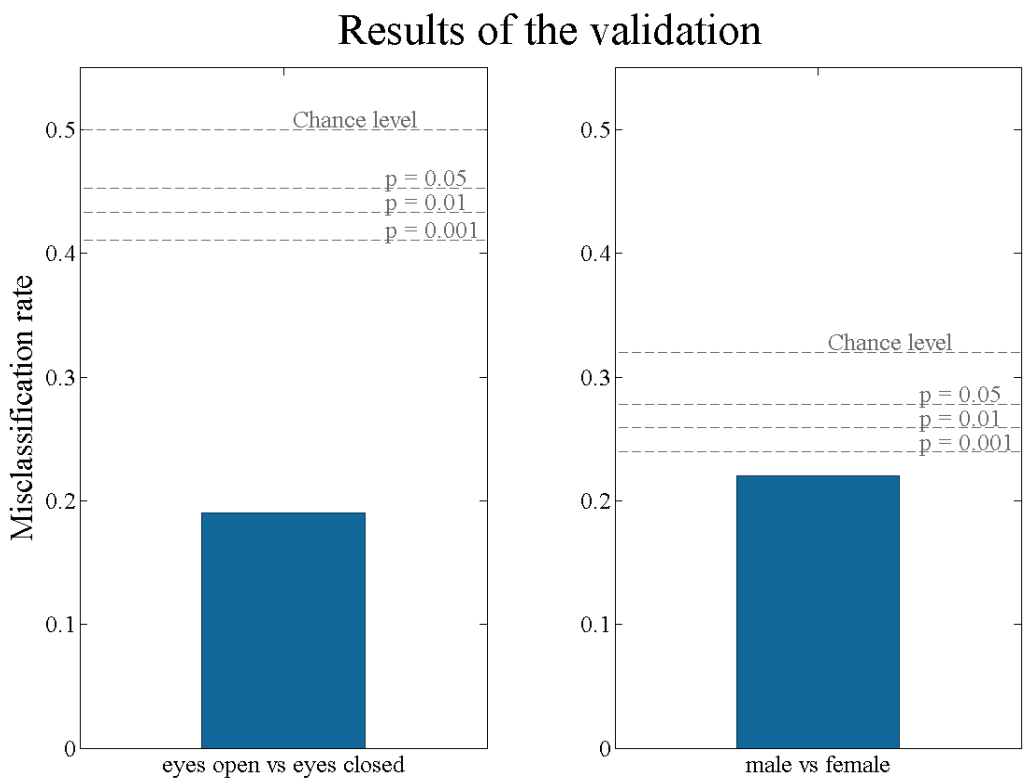

**Figure 3**: Results of the validation. Both cases have 309 subjects but note that the chance levels and significance levels of the two panels are different as we had 209 females and 100 males.

These results indicate that the failure to predict personality traits from spectral components of resting state EEG is probably not due to systematic algorithmic or implementation mistakes.

## 3. Discussion

In the present study we asked whether it is possible to predict the Big Five personality traits from resting state EEG data. Previous studies investigating the neurobiological correlates of personality have focused on single parameters such as the power of specific frequency components. Here we approached the research question from the data-driven perspective, using machine learning to assess how much information about personality traits is contained in the spectral dynamics of the resting state EEG. In particular, we analyzed the data from the complete power spectrum and all the 32 electrodes. We used nested cross-validation on 309 subjects to estimate the accuracy of the classifiers.

Our results indicated that it is not possible to predict the Big Five personality traits from the power spectra of resting state EEG data. We furthermore showed that our classifiers are also not successful in predicting the 10 lower-order aspects of Big Five. Although for *Politeness* the *p* value was lower than 0.05, the result did not survive a correction for multiple comparisons. Finally, we demonstrated that the inability to classify the personality traits was not caused by our analysis pipeline. Using the same methods, we were able to differentiate eyes open recordings from eyes closed ones with a misclassification rate of 19%. Furthermore, we could decode whether the subject was male or female with a misclassification rate of 22%.

Given the comprehensive statistical approach used in this study, the results, despite being negative, have implications for personality neuroscience. On the one hand, the limitations of the recording and analysis techniques used here can be informative. Our study relied on the power spectrum of each channel as the basis for selecting features or explanatory variables that may predict personality traits. Spectral rather than temporal characteristics of resting state EEG were explored as all points in time are statistically equivalent in this type of signal (as opposed to evoked or induced potentials). Nevertheless, it is possible that features not captured by power such as oscillatory phase or temporal correlations in channel or source space might add some extra information. For instance, some studies have found relationships between trait measures such as Neuroticism and negative correlations between amplitudes of higher (alpha/beta) and lower (theta/delta) frequencies (Schutter & Knyazev, 2012). Note however that in many of these studies, power levels in certain frequencies are also related to the cross-frequency coupling indices as well as to the relevant trait measures (e.g. Knyazev et al., 2003). It thus remains to be seen if different feature extraction strategies would increase the success of classifying personality scores from resting state EEG. In general, one must notice that although our sample size is large in the context of EEG measures, machine learning techniques are typically applied to larger data sets. Also, any machine learning approach might suffer from several trade-offs and arbitrariness at each of its different stages (data preprocessing, feature extraction, feature selection, model selection, model training, validation).

Another technical explanation for the current findings might be the limited nature of EEG as a measure of brain processes. EEG is sensitive to only a subset of electrical events in the brain, probably reflecting the synchronized local field potentials of suitably aligned cortical pyramidal neurons (Lopes da Silva, 2013). There is some evidence suggesting that the trait-relevant differences may instead be found on the level of brain structure (e.g. DeYoung et al., 2010) or activation in subcortical areas that are unlikely to contribute directly to scalp EEG (e.g. Cunningham, Arbuckle, Jahn, Mowrer, & Abduljalil, 2010).

These methodological considerations notwithstanding, the present findings may also have conceptual implications. Individual differences in brain processes can either be stable dispositions evident in majority of situations (i.e. situation-independent traits) or characteristic responses to specific stimuli (situation-dependent traits; Fleeson & Noftle, 2009; Mischel & Shoda, 1998; Stemmler & Wacker, 2010). Given the lack of stimulation, the resting state measurement is optimized for discovering the former rather than the latter type of individual differences. In this framework, the failure to relate resting state EEG to Big Five personality traits documented here suggests that the brain substrate of personality might involve situation-dependent responsiveness rather than differences in baseline activity.

A relevant example can be found from the literature relating anterior EEG asymmetry to personality. Although many researchers implicitly assume the resting condition to be optimal for quantifying trait asymmetry, there is evidence that trait-related changes in asymmetries are best captured in response to some relevant stimulation (Coan, Allen, & McKnight, 2006). Furthermore, there is evidence that the trait-asymmetry correlations that emerge in the resting state data, may also be driven by the situational features of the resting measurement occasion. For instance, anterior asymmetry responses distinguished male participants based on their trait Defensiveness only at the presence of an attractive female experimenter (e.g. Kline, Blackhart, & Joiner, 2002). Given that we analyzed data from different studies collected by different experimenters, such situational factors should have fairly randomly distributed effects within the present data. This reduces the risk of situation-mediated EEG-personality covariance being registered as evidence for correlations on the trait level.

In summary, regarding the discovery of neural bases of personality traits, the present null-finding may constitute a false negative in the sense that technical limitations of EEG recording and/or the employed analysis techniques precluded us from detecting true trait differences in brain activity. In addition, even while the size of the current sample (309 participants) is similar to some normative datasets, we cannot exclude the possibility that more data would have provided significant results. On the other hand however, the results might imply that the brain substrate of personality may exist on the level of characteristic responsiveness rather than baseline activity, in line with several findings in modern personality neuroscience (Stemmler & Wacker, 2010).

## 4. Methods

### 4.1 Sample and procedure

The sample consists of 309 participants of 12 different cognitive EEG experiments conducted at the University of Tartu (100 males; age range 18-42, $M = 21.9$, $\sigma = 3.4$). All participants were volunteers recruited among university students and the general population. All measurements were approved by the Ethics Review Board of Tartu University. Only subjects who completed a personality questionnaire and had at least 50% of the originally recorded EEG data retained after artifact rejection were included in the sample.

The resting state EEG data analyzed here were collected prior to all other experimental tasks. Two different measurement protocols were used. In 5 experiments the resting state signal was recorded in two contiguous sections – one with eyes open and the other with eyes closed. Each section lasted either for 1 minute (in one experiment with 84 participants) or 2 minutes (in four experiments with 94 participants). In the remaining 7 experiments (131 participants) 3 separate 1-minute measurements with eyes open and eyes closed were interleaved resulting in 3 minutes in total for both the eyes open and eyes closed conditions.

The measurements took place in a dimly lit and quiet room. Participants sat in a comfortable office chair 1 or 1.15 m away from a computer screen. They were instructed to relax and avoid excessive body and eye movements. During the eyes open condition they were also required to fixate on a black cross in the middle of a grey screen. After instructions, participants remained alone in the room during the actual recording.

## 4.2 EEG recording and preprocessing

A BioSemi ActiveTwo (BioSemi, Amsterdam, Netherlands) active electrode system was used to record signals from 32 scalp locations; two reference electrodes placed on earlobes and four ocular electrodes (above and below the left eye and near the outer canthi of both eyes). The data were recorded with 0.16 – 100 Hz band-pass filter and 1024 or 512 Hz sampling rate.

Offline pre-processing was implemented in Matlab (MathWorks, USA) and EEGLAB (Delorme & Makeig, 2004) software. The data were re-referenced to digitally linked earlobes and resampled to 512 Hz as necessary. Eye-movement artifacts were corrected using Independent Component Analysis (ICA). Infomax ICA algorithm was trained on a separate copy of data that was first high-pass filtered (half-amplitude cut-off at 1 Hz) and then cleaned of large artifacts by screening 1-second epochs for excessive muscular noise (EEGLAB rejspec 15-30 Hz; +/-45 dB). If more than 2% of epochs were marked for rejection based on a single channel, the channel was removed before rejecting the remaining epochs with artifacts. Independent components capturing eye-blinks as well as vertical and horizontal eye-movements were visually identified and removed before reconstructing the whole duration of unfiltered and unsegmented data (Debener, Thorne, Schneider, & Viola, 2010). The ICA-pruned continuous data were cut into 4-second epochs and the mean voltage of each epoch was removed as a baseline. All segments where voltage fluctuations exceeded 200 µV were marked as artifacts. If this criterion was violated in only a single channel for more than 2% or the trials, this channel was removed before removing the remaining segments with artifacts. All rejected channels were spherically interpolated (EEGLAB *eeg_interp*).

The power spectral density values were computed using the Fourier transform applied separately to data from each channel and retained segment. The power estimates between 0.25 and 95 Hz were added and divided by the number of available trials within eyes closed and eyes open condition.

## 4.3 Personality measures

Prior to visiting the lab, participants filled in a personality questionnaire in a dedicated online environment (kaemus.psych.ut.ee). All 309 participants completed the EE.PIP-NEO inventory which uses 240 items to assess the 5 dimensions (see figure 4) as well as 30 facets of the FFM (Mõttus, Pullmann, & Allik, 2006).

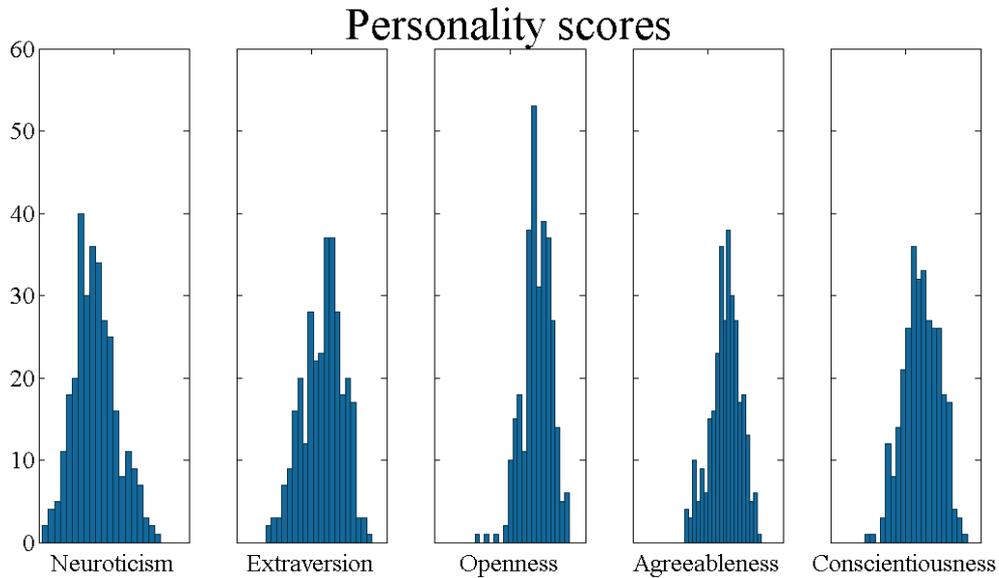

**Figure 4**: Histograms of all 5 personality dimension scores. Note that the distributions are rather normal and personality score units on the x-axes are arbitrary.

Scores for the 10 lower-order aspects of the Big Five were computed as averages of the relevant facet scores of the FFM (Judge, Rodell, Klinger, Simon, & Crawford, 2013):

- Volatility = N2, N5
- Withdrawal = N1, N3, N4, N6
- Enthusiasm = E1, E2, E5, E6
- Assertiveness = E3, E4, E5
- Intellect = O5
- Openness = O1, O2, O3, O4, O6
- Compassion = A1, A3, A6
- Politeness = A2, A4, A5
- Industriousness = C1, C4, C5
- Orderliness = C2, C3, C6

Median split was performed on each personality trait, resulting in two equal size classes "high" and "low" score for each personality trait.

### 4.4 Classification of the personality traits

We used a nested cross-validation approach. For predicting an out of sample subject we used 10-fold cross-validation. In order to choose the best hyper-parameters we used again 10-fold cross-validation for 90% of the data. After finding the best combination of hyper-parameters we re-trained it on the same 90% of the data. Each of the five personality traits was treated separately. Matlab code for the nested cross-validation can be found from GitHub repository: https://github.com/kristjankorjus/PredictingPersonalityFromEEG

### 4.4.1 Nested Cross-Validation

In order to find the best hyper-parameters 10-fold cross-validation was used on the 90% of the data inside of cross-validation. The full list of hyper- parameters is specified here:

- different data types: using power spectra of eyes open data or eyes closed data (*2 options*)
- pooling of electrodes: using all electrodes separately or taking regions of interests (left frontal (F3, F7, AF3, Fp1); right frontal (F4, F8, AF4, Fp2); mid-frontal (Fz, Cz, FC1, FC2); left central (FC5, CP5, T7, P7, C3); right central (FC6, CP6, T8, P8, C4); mid-parietal (Pz, CP1, CP2, P3, P4); and occipital (PO3, PO4, O1, Oz, O2)) (*2 options*)
- pooling of frequencies: using all frequencies, using customized pooling for frequencies (taking information from spectrum averaged over all subjects and channels such that from 0 Hz to 14.75 Hz bands of 0.25 Hz were used; from 15 Hz to 24.75 Hz bands of 0.5 Hz; from 25 Hz to 36.75 Hz bands of 1 Hz; from 37 Hz to 95 Hz bands of 10 Hz were used) or using 11 bands based on the literature (0.5 to 1.25, 1.5 to 4.25, 4.5 to 8.25, 8.5 to 12.25, 12.5 to 20.25, 20.5 to 30.25, 30.5 to 40.25, 40.5 to 50, 50.25 to 69.75, 70 to 89.75, 90 to 95, all values in Hz) (*3 options*)
- normalization of the data: using the data without normalization, normalizing each row (each subject) by taking the z-score (subtracting the mean, dividing by standard deviation) or normalizing each column (each feature) (*3 options*)
- for reduction of dimensionality of the data, principal component analysis was used with different amount of total variance explained by principal components: 90% or 70% of variance explained. Or no principal component analysis was used at all (*3 options*)
- value of the box constraint C for the soft margin of Support-Vector-Machine (SVM): 0.01 or 100 (*2 options*)
- type of kernel for SVM: no kernel (linear SVM) or a radial basis function (RBF) kernel (non-linear SVM) were used. In addition, the sigma parameter controlling the width of the RBF was selected among two possible values using randomized data as explained in the next paragraph (*3 options in total*).

To find a suitable sigma parameter of RBF for our dataset we estimated how the over-fitting error for a known case depended on this parameter. In particular, we first randomized the classes of the dataset, and let the classifier train and predict the full data. For small values of the sigma, the classification error will tend to 0 indicating over-fitting – every neighborhood around a sample is classified with the label of the contained sample, and therefore lacking of any generalization power. As the sigma increased, the error rate started to increase. Sigma was fixed when the error rate reached 0.1 or 0.3 as these were optimistic prior beliefs about the final classification error.

In total, these 648 combinations of classifier hyper-parameters were explored for each personality trait. For the whole cross-validation phase, partitioning of data was fixed to make different hyper-parameters more comparable. The smallest misclassification error for each personality trait determined the set of hyper-parameters, which were used in testing.

### 4.4.2 Cross-Validation

In order to estimate the misclassification error rate, each personality trait of 309 subjects was predicted with the best hyper-parameters found and model trained in the nested cross-validation phase.

Statistical significance was estimated using binomial test with the null hypothesis that two categories are equally likely to occur.

**4.5 Analyzing continuous data**

Continuous scores were first normalized (mean was subtracted and divided by standard deviation). Instead of a SVM classifier, LASSO and elastic net regressions were used (Matlab function *lasso*). All hyper-parameters which were not related to the specifics of SVM classifiers, were scanned from the same range as described in the Cross-Validation section above. In addition, the *alpha* parameter was scanned with three options: 1, 0.5 and 0.01. Parameter *alpha* = 1 represents LASSO regression, a*lpha* close to 0 approaches ridge regression, and the 0.5 represents elastic net optimization. For the regularization parameter, we used the recommended *lambda* such that MSE is within one standard error of the minimum (see *Lambda1SE* in the *lasso* documentation in MATLAB).

In the continuous case, the null hypothesis was that the best prediction for each personality trait, which minimizes the mean squared error, is the mean of the personality score. Statistical significance analysis was performed using a permutation test: scores were sampled with replacements from the score distribution.

**4.6 Testing the influence of age and gender related variability**

To test possible age and gender related systematic personality variability, all trait scores were normalized in relation to age- and gender-specific means and standard deviations ($S_{\text{norm}} = \frac{S_{\text{raw}} - \mu_{\text{reference}}}{\sigma_{\text{reference}}} \cdot 10 + 50$). The reference data were obtained from a normative sample of the EE.PIP-NEO ($n = 1564$; 889 males; age range 16-86; $M = 15.8, \sigma = 12.1$). Based on the age and gender of the participant, one of 20 reference groups were selected (9 age brackets between 15 and 59 with 5 year steps and one bracket for 60 and above). For 53 participants of the present sample for whom age data were unavailable, the sample mean age was used for reference group identification. Again, median split and original pipeline described above was used.

**4.7 Assessment of 10 lower-order aspects**

Instead of five personality traits, each trait has a natural subdivision into two:

- neuroticism: *volatility* and *withdrawal*
- extroversion: *assertiveness* and *enthusiasm*
- openness to experience: *intellect* and *openness*
- agreeableness: *compassion* and *politeness*
- conscientiousness: *industriousness* and *orderliness*

All of the 10 sub-traits were classified using the original above-described pipeline.

**4.8 Classification of eyes closed vs eyes opened data**

To test the validity of the whole pipeline, classification was performed in a situation with clear pattern of information present. In particular, the data were divided into two sets such that half of the subjects had eyes open data and another half had eyes closed, in total two classes. Notice that no subject appears in two classes. Thus we still have 309 subjects and two classes but for each subject the data is taken from either eyes open or eyes closed condition. Again, original pipeline was used.

**4.8 Classification of male vs female subjects**

For another test of the validity of the whole pipeline, classification was performed also in a situation with likely pattern of information present. In particular, the gender of subjects was used: 209 female subjects and 100 male subjects were used. Notice that the chance misclassification level is 32.4%. Again, original pipeline was used.

# 5. Author Contributions

Kristjan, Andero, Raul and Jaan analyzed the data. Andero, Helen, Nele, Kairi and Jüri organised the data collection. Kristjan, Andero, Raul and Jaan wrote the manuscript with input from all other authors.

# 6. Conflict of Interest Statement

The authors declare that the research was conducted in the absence of any commercial or financial relationships that could be construed as a potential conflict of interest.

# 7. Acknowledgments

This research was supported by the European Regional Development Fund through the Estonian Center of Excellence in Computer Science, EXCS. K.K., R.V., and J.A. also thank the financial support from the Estonian Research Council through the personal research grants P.U.T. program (PUT438 grant). JA was also supported by Estonian Research Council grant IUT20-40. The work by A.U., H.U., N.K., K.K and J.A. was supported by a Primus grant (#3-8.2/60) from the European Social Fund to Anu Realo and grants from Estonian Ministry of Science and Education (SF0180029s08 and IUT02-13). H.U. and N.K. are members of the Doctoral School of Behavioural, Social and Health Sciences created under the auspices of the European Social Fund. The data collection was also supported by the Estonian Science Foundation (grant #8332). We also thank our students who helped in collecting the data.